\documentclass[aps,amsmath,showpacs,amsfonts,10pt]{revtex4}
\usepackage{epsfig,graphicx}
\usepackage[english]{babel}
\usepackage{amsfonts}
\usepackage{amsmath}
\usepackage{latexsym}
\usepackage{graphics,bm}
\usepackage{dcolumn}
\usepackage{bm}
\usepackage{rotating}

\begin{document}

\title{Influence of Virtual Photon Process on the Generation of Squeezed Light from Atoms in an Optical Cavity}

\author{Aranya B Bhattacherjee}
\affiliation{School of Physical Sciences, Jawaharlal Nehru University, New Delhi-110067, India }

\begin{abstract}
We show that a collection of two-level atoms in an optical cavity beyond the rotating wave approximation and in the dispersive and strong-coupling regime constitutes a nonlinear medium and is capable of generating squeezed state of light. It is found that squeezing produced in the strong-coupling regime is significantly higher compared to that produced in the dispersive limit. On the other hand, we also show that it could be possible to observe the Dicke superradiant quantum phase transition in the dispersive regime where the $\vec{A}^{2}$ term is negligible.  Such a system can be an essential component of a larger quantum-communication system.
\end{abstract}


\maketitle

\section{Introduction}

Non-classical states of light sources are one of the essential component for implementing quantum-information processing and secure telecommunication quantum systems\citep{1,2,3,4}. The experimental generation of quantum light have made remarkable progress in the recent years, thereby opening the path for implementing novel quantum devices based on quantum light\citep{5,6,7,8,9,10,11,12,13,14,15,16}. A paradigmatic example of quantum light source is the optical parametric amplifier, which is usually realized with nonlinear crystals in resonators\citep{17}. In the recent past there have been theoretical proposals that a single atom and two atoms in a suitable setup can act as a nonlinear medium to generate squeezed light \citep{18,19,20,21,22,23}. More recently, it was proposed that a Bose-Einstein condensate in an optical cavity also generate squeezed light with the two-body atom-atom interaction appearing as a coherent handle to control this generation \citep{24}.
In this paper, we study the generation of squeezed states of light from a medium comprising of $N$ two-level atoms confined in an optical cavity in the non-rotating wave approximation (non-RWA). The fully quantum-mechanical model for the interaction of light and matter was given by Jaynes and Cummings \citep{25}.

\begin{equation}
H_{JCM}=\frac{\hbar \omega_{eg}}{2}\sigma^{z}+\hbar \omega_{c} a^{\dagger}a+\hbar g \sigma^{x} (a^{\dagger}+a),
\end{equation}

where $\hbar \omega_{eg}$ is the level spacing between the ground and the excited state of the atoms, $\omega_{c}$ is the frequency of the electromagnetic field mode, and $g$ is the dipole interaction strength. $\sigma^{\alpha}$, $\alpha=x,y,z$ are the Pauli matrices while $a^{\dagger}$ and $a$ denote the bosonic creation and annihilation operators of the electromagnetic field mode. In the interaction picture the co-rotating terms in the Jaynes-Cummings model (JCM), oscillate with the phase factors $e^{\pm i (\omega_{eg}-\omega_{c})t}$, while the counter-rotating terms oscillate with the phase factors $e^{\pm i (\omega_{eg}+\omega_{c})t}$. Near resonance, the detuning $(\omega_{eg}-\omega_{c})$ is small, $|\omega_{eg}-\omega_{c}|<<\omega_{eg}+\omega_{c}$ and hence the co-rotating terms oscillate slowly, whereas the counter-rotating terms oscillate fast.In addition, if the atom-field coupling $g_{o}$ is sufficiently weak, one can then replace the counter-rotating terms by their vanishing time average. This leads to the usual RWA-JCM.

Thus the RWA-JCM is valid for $g_{o}$ $<<$ $|\omega_{eg}-\omega_{c}|$ $<<$ $\omega_{eg}+\omega_{c}$, where the first inequality is the dispersive limit \citep{26}. Now in the adiabatic limit i.e. $\omega_{eg}<< \omega_{c}$ or $\omega_{eg}>> \omega_{c}$, the inequality  $|\omega_{eg}-\omega_{c}|<<\omega_{eg}+\omega_{c}$ is no longer valid and hence RWA breaks down. In fact in the adiabatic limit $|\omega_{eg}-\omega_{c}|\approx (\omega_{eg}+\omega_{c})$ and hence the counter-rotating terms are equally significant.

In the current paper, we will work in the dispersive-adiabatic limit i.e. only the inequality $g_{o}<< |\omega_{eg}-\omega_{c}|$ is satisfied and compare our results with those in the strong-coupling-adiabatic limit (i.e when $g_{o}>> |\omega_{eg}-\omega_{c}|$, $g_{0}>>\gamma $ (cavity field decay rate) and $|\omega_{eg}-\omega_{c}|\approx (\omega_{eg}+\omega_{c})$). The strong coupling between the atom and the field which is difficult to realize experimentally also does not allow us to ignore the counter-rotating terms. The importance of the dispersive limit lies in the use of two-level systems to simulate quantum spin chains or building quantum switch.  The importance of the strong-coupling regime is evident from the study of the Dicke model which shows strong atom-field entanglement \citep{27}. In recent years, the strong-coupling regime in atomic systems has been achieved \citep{28}. The condition $g_{o}<< |\omega_{eg}-\omega_{c}|$ is then a less stringent condition compared to the strong-coupling condition.

\section{The Model}

\begin{figure}[h]
\hspace{-0.0cm}
\includegraphics [scale=0.55] {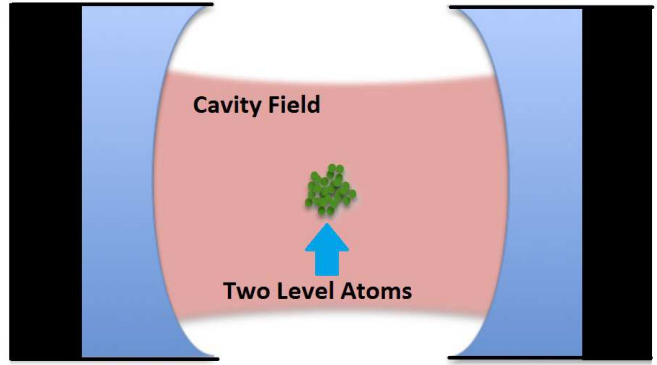}
\caption{Sketch of the system considered in this paper: Collection of $N$ identical atoms confined in a small volume at the center of the optical cavity coupled to the quantum field of the cavity by their electric dipole.}
\label{f1}
\end{figure}

Let us consider a collection of $N$ identical, non-interacting atoms coupled to the quantum field of a cavity resonator by electric dipole interaction (Figure 1). The bare Hamiltonian for a single $j^{th}$ atom is

\begin{equation}
H_{j}=\sum_{i=1}^{M}\frac{P_{i_{j}}^{2}}{2 m_{i_{j}}}+V_{j},
\end{equation}

where $V_{j}$ is the position dependent generic potential energy. The index $i_{j}$ denotes one of the $M$ particles of charge $q_{i}$ making up the atom. The momentum and mass of the $j^{th}$ atom is $\sum_{i} P_{i_{j}}$ and $\sum_{i} m_{i_{j}}$. The Hamiltonian for $N$ atoms is $H_{a}=\sum_{j=1}^{N}H_{j}$. Now, if we consider the interaction of the $N$-atoms with the quantum field in the optical cavity, then according to the principle of minimal coupling, $\vec{P}_{i_{j}}\rightarrow \vec{P}_{i_{j}}-q_{i_{j}}\vec{A}(\vec{r}_{i})$, where $\vec{A}$ is the vector potential field in the region occupied by the atoms.Considering only one cavity mode and neglecting the spatial variation of the field in the region occupied by the atoms, we can write $\vec{A}(\vec{r})=\vec{A_{o}}(a^{\dagger}+a)$. Here $a^{\dagger}(a)$ is the cavity mode creation and annihilation operator. Therefore, the Hamiltonian for the interaction part is ,

\begin{equation}
H_{int}=-\sum_{j=1}^{N} \sum_{i=1}^{M} \left( \frac{q_{i_{j}}}{m_{i_{j}}} \vec{P}_{i_{j}}\bullet \vec{A}_{o} (a^{\dagger}+a) + \frac{q_{i_{j}}^{2}}{2 m_{i_{j}} } A_{o}^{2} (a^{\dagger}+a)^{2}\right )
\end{equation}

The $H_{int}$ can be written as

\begin{equation}
H_{int}=-i \hbar \Omega_{o} (a^{\dagger}+a) b^{\dagger}+h.c +\hbar D (a^{\dagger}+a)^{2},
\end{equation}

where $\Omega_{o}=\frac{\omega_{eg}}{\hbar}\vec{d}_{eg}\bullet \vec{A}_{o} \sqrt{N}$. $\omega_{eg}$ is the atomic transition frequency while $\vec{d}_{eg}$ is the electric dipole matrix element. Also $D=\sum_{i=1}^{M} \frac{q_{i}^{2}}{2 m_{i}}\frac{N A_{o}^{2}}{\hbar}$.
The bosonic operator $b^{\dagger}$ is the bright excitation operator defined as

\begin{equation}
b^{\dagger}=\frac{1}{\sqrt{N}} \sum_{j=1}^{N} \left ( |e> <g|\right ),
\end{equation}

where $|e>$ and $|g>$ are the two eigenstates of each atom. The atomic Hamiltonian can be rewritten as $H_{a}=\hbar \omega_{eg} b^{\dagger}b$, while the cavity mode Hamiltonian is written as $H_{cav}=\hbar \omega_{c} a^{\dagger}a$. The total Hamiltonian is $H=H_{a}+H_{cav}+H_{int}$.

\section{Squeezing of the intra-cavity field}

The Heisenberg equations of motion for the atomic and photonic operator is

\begin{equation}
\dot{a}=-i (\omega_{c}+2 D)a-i 2D a^{\dagger}+\Omega_{o}(b-b^{\dagger})-\gamma a,
\end{equation}

\begin{equation}
\dot{b}=-i \omega_{eg}b-\Omega_{o}(a+a^{\dagger}).
\end{equation}

In the limit of large atom-photon detuning, we adiabatically eliminate the atomic degrees of freedom, $b=\frac{i\Omega_{o}(a+a^{\dagger})}{\omega_{eg}}$. In cavity QED systems, the collective coupling $U_{o}= \Omega_{o}^{2}/\omega_{eg}$ and $D$ are dependent on each other. Infact $D=\alpha U_{o}$, where $\alpha$ is a parameter. The value of $\alpha$ is determined by the Thomas-Reiche-Kuhn sum rule \citep{29,30}. It is important to note that the $\vec{A}^{2}$ term is significant only in the strong-coupling limit. This yields the following Heisenberg-Langevin equation for the field operators $a$ and $a^{\dagger}$,

\begin{equation}
\dot{a}=-\left( i[\omega_{c}+2 U_{o} (\alpha-1)]+\gamma \right)a-2 i U_{o} (\alpha-1)a^{\dagger}+F(t),
\end{equation}

\begin{equation}
\dot{a}^{\dagger}=-\left( -i[\omega_{c}+2 U_{o} (\alpha-1)]+\gamma \right)a^{\dagger} +2 i U_{o} (\alpha-1)a + F^{\dagger}(t).
\end{equation}

Here $\gamma$ represents the cavity decay and $F(t)$ is the associated noise operator with the following properties: $<F(t)>=0$, $<F(t) F(t')>=<F^{\dagger}(t) F(t')>=<F^{\dagger}(t) F^{\dagger}(t')>=0$ and $<F(t) F^{\dagger}(t')>=2 \gamma \delta(t-t')$. We see that the cavity frequency is shifted as $\omega_{c}\rightarrow \omega_{c} + 2U_{0}(\alpha-1)$ due to atomic back-action and $\vec{A}^{2}$ term.

In the steady state, from Eqns.(8) and (9), one can calculate $|a_{s}|^2$ (intra-cavity steady state photon number) as,

\begin{equation}
|a_{s}|^{2}=-\frac{\left( \gamma^{2}+(\omega_{c}+2U_{0}(\alpha-1))^{2}\right) }{\left( \omega_{c}^{2}+4 \omega_{c} U_{0}(\alpha-1)+\gamma^{2} \right)},
\end{equation}

In the dispersive regime, one can put $\alpha=0$ since the atom-field coupling is weak and $|a_{s}|^{2}\rightarrow \infty$ at the critical atom-photon coupling $\Omega_{c}=\sqrt{\frac{\omega_{eg}(\omega_{c}^{2}+\gamma^{2})}{4 \omega_{c}}}$. This is the Dicke superradiant quantum phase transition (QPT) \citep{27}. In the strong coupling regime, $\alpha$ is finite and $\alpha>1$ and hence the NO-GO theorem is valid \citep{29,30}. In the case of systems comprising of two-level atoms in a cavity coupled to the electromagnetic field of the cavity through their electric dipole, the inclusion of the $\vec{A}^{2}$ term in the Dicke Hamiltonian forbids the occurrence of the QPT as a consequence of the Thomas-Reiche-Kuhn sum rule for the oscillator strength.  This analysis indicates that it could be possible to observe the Dicke superradiant QPT in a collection of two level atoms in an optical cavity in the dispersive regime where the $\vec{A}^{2}$ term responsible for the NO-GO theorem is negligible. Indeed, a superradiant phase transition does not require $\omega_{eg}\approx \omega_{c}$ \citep{29}.

In order to study the squeezing properties of the signal mode in the steady state, we have to evaluate $<a>$, $<a^{2}>$,$<a^{\dagger 2}>$ and $<a^{\dagger}a>$. Taking into account that all correlation functions involving the noise operators are zero except $<F a^{\dagger}>=<a F^{\dagger}>= \gamma$, yields the following steady state values,

\begin{equation}
A_{1}=<a^{2}>_{ss}=\frac{\Omega_{P} \Gamma^{*}}{2(|\Omega_{P}|^2-\Gamma \Gamma^{*})},
\end{equation}

\begin{equation}
A_{2}=<a a^{\dagger}+ a^{\dagger}a>_{ss}=\frac{- \Gamma \Gamma^{*}}{(|\Omega_{P}|^2-\Gamma \Gamma^{*})},
\end{equation}

\begin{equation}
A_{3}=<a^{\dagger 2}>_{ss}=\frac{\Omega^{*} _{P} \Gamma}{2(|\Omega_{P}|^2-\Gamma \Gamma^{*})},
\end{equation}

where $\Omega_{P}=-2i U_{o}(\alpha-1)$ and $\Gamma=-(i[\omega_{c}+2 U_{o}(\alpha-1)]+\gamma)$. To calculate the variances, the field is expressed in terms of Hermitian operators $X_{1}=\frac{1}{2} (a e^{-i \theta/2}+ a^{\dagger} e^{i \theta/2})$ and $X_{2}=\frac{1}{2i} (a e^{-i \theta/2}- a^{\dagger} e^{i \theta/2})$.
The variances of these operators in the steady state for $\theta=0$ are,

\begin{equation}
(\Delta X_{1}^{2})_{ss}=\frac{1}{4} \frac{\gamma^2+\omega_{c}^{2}+2 \omega_{c} U_{o} (\alpha-1)}{\gamma^2+\omega_{c}^{2}+4 \omega_{c} U_{o} (\alpha-1)},
\end{equation}

\begin{equation}
(\Delta X_{2}^{2})_{ss}=\frac{1}{4} \frac{\gamma^2+\omega_{c}^{2}+8 U_{o}^{2}(\alpha-1)^{2}+6 \omega_{c} U_{o}(\alpha-1)}{\gamma^2+\omega_{c}^{2}+4 \omega_{c} U_{o} (\alpha-1)},
\end{equation}

\begin{figure}[h]
\hspace{-0.0cm}
\begin{tabular}{cc}
\includegraphics [scale=0.55]{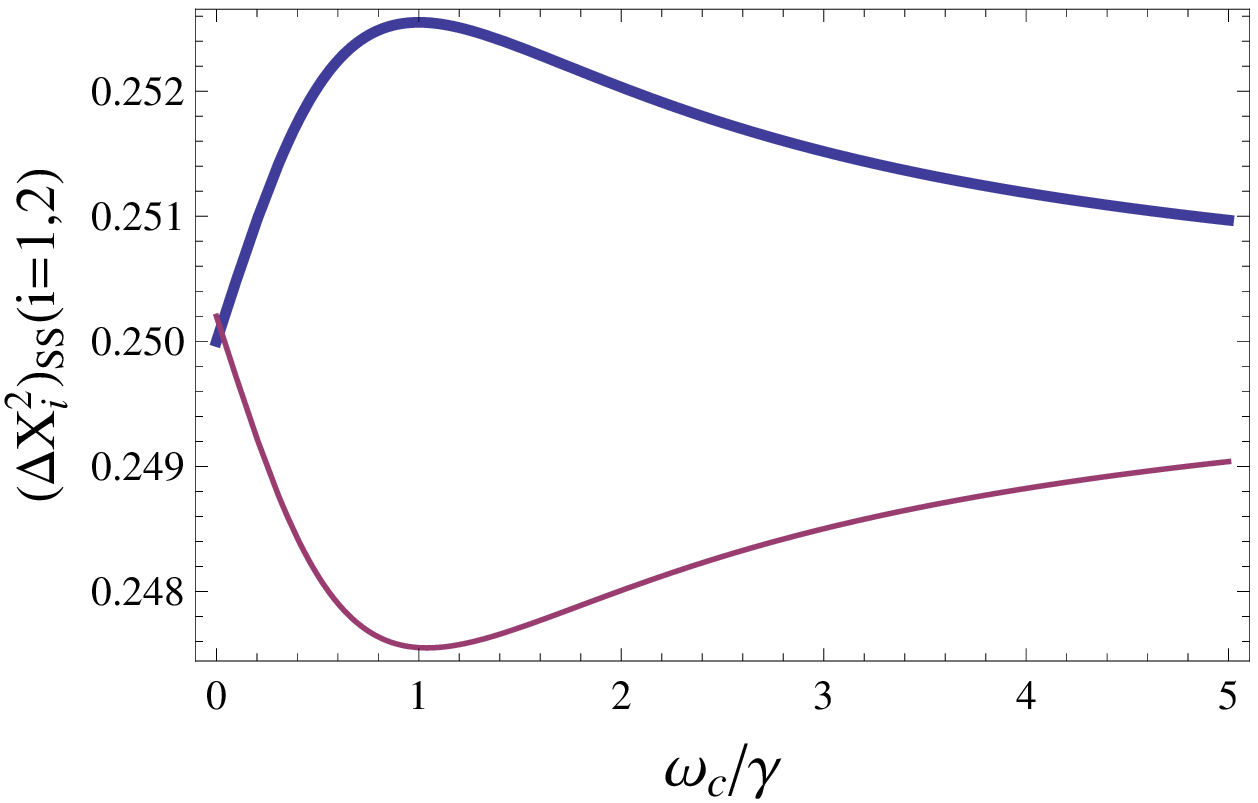} \includegraphics [scale=0.55] {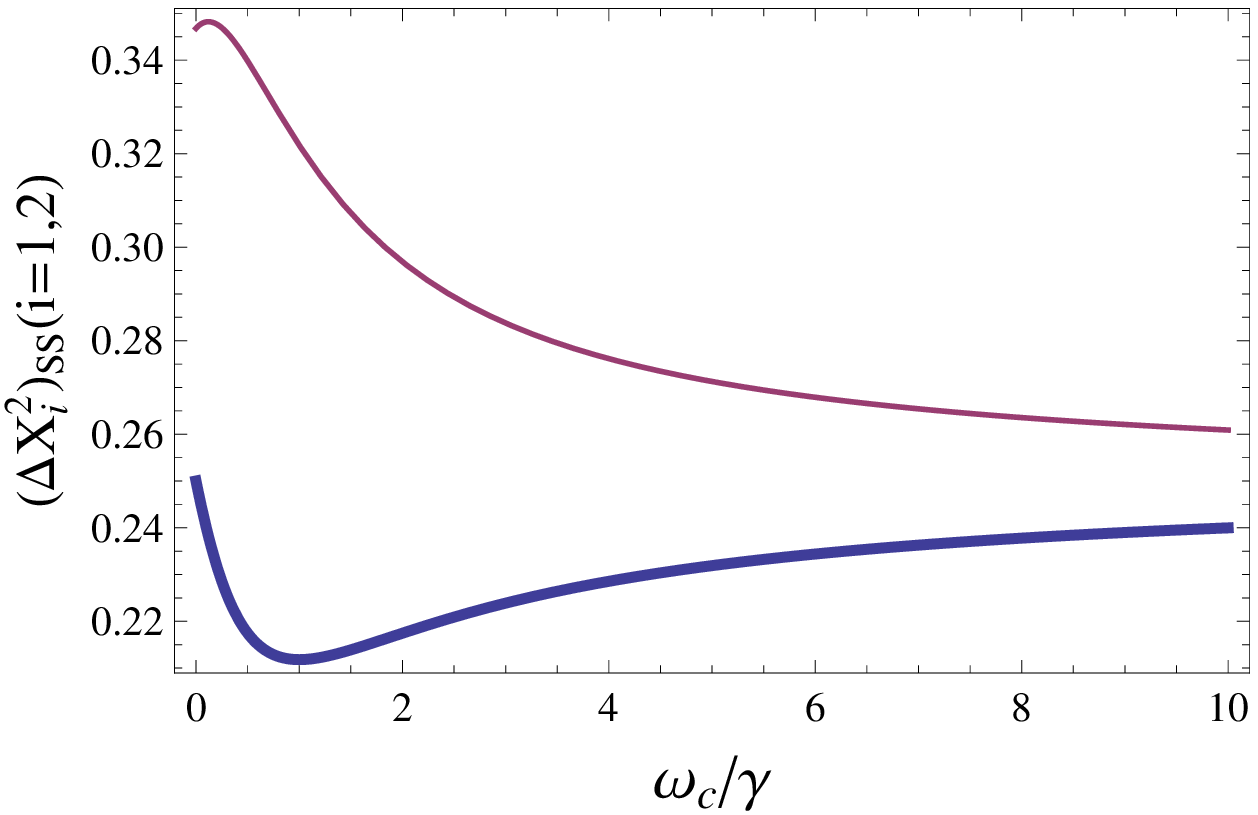}\\
 \end{tabular}
\caption{Left Plot: Plot of the variances versus $\omega_{c}/\gamma$ for $U_{0}=0.01 \gamma$ and $\alpha=0$ . Right Plot: $U_{0}=1.1 \gamma$ and $\alpha=1.2$. $\Delta X_{1}^{2}$ (thick line) and $\Delta X_{2}^{2}$ (thin line)}
\label{f2}
\end{figure}

Note that to be in the dispersive-adiabatic limit, we keep $Uo<< \omega_{c}$. In RWA limit, $(\Delta X_{1}^{2})_{ss}=(\Delta X_{1}^{2})_{ss}=1/4$ i.e. both the quadratures are in the coherent state. Figure 2 illustrates the quadrature variances in the dispersive-adiabatic limit (left plot) and the strong-coupling-adiabatic limit (right plot). In the dispersive limit since $g_{0}<< |\omega_{eg}-\omega_{c}|$, we take $\alpha=0$. Evidently, squeezing of the cavity field is moderately enhanced for the strong-coupling case compared to that in the dispersive case. The quadrature variances does not reflect the true experimental situation which is concerned with the field outside the cavity. We address this issue in the next section.

\section{Squeezing of the output field}

The Heisenberg-Langevin equations of motion for the cavity field including the input field is written as,

\begin{equation}
\dot{a}=\Gamma a + \Omega_{P} a^{\dagger}+ \sqrt{a \gamma} a_{in},
\end{equation}

\begin{equation}
\dot{a}^{\dagger}=\Gamma^{*} a^{\dagger} + \Omega_{P}^{*} a^{\dagger}+ \sqrt{a \gamma} a^{\dagger}_{in},
\end{equation}

where $a_{in} (a^{\dagger}_{in})$ is the destruction(creation) operator for the input cavity field. The relationship that connects the external fields (both input and output) and the intracavity field is \citep{17},

\begin{equation}
a_{out}(t)+a_{in}(t)=\sqrt{2 \gamma} a(t),
\end{equation}

where $a_{out} (a^{\dagger}_{out})$ is the destruction(creation) operator for the output cavity field. Eliminating the internal cavity mode using Eqns.(16)-(17) in the Fourier space, we obtain

\begin{equation}
a_{out}=\frac{(\omega^{2}-\omega_{c}\Delta_{c}-\gamma^{2}-2i\omega \gamma)}{(\omega^{2}-\omega_{c}\Delta_{c}-\gamma^{2})^{2}+4 \omega^{4} \gamma^{2}}\left( (-\omega^{2}+\omega_{c}\Delta_{c}-\gamma^{2}+2i\gamma \tilde{\Delta}_{c})a_{in}-2 \gamma \Omega_{P} a^{\dagger}_{in}\right)
\end{equation}

where ${\Delta}_{c}=\omega_{c}+4 U_{o}(\alpha-1)$ and $\tilde{\Delta}_{c}=\omega_{c}+2 U_{o}(\alpha-1)$. The squeezing spectrum for the two quadratures is then obtained as \citep{17},

\begin{equation}
S_{1out}(\omega)=\frac{1}{4} \frac{(\omega^{2}-\omega_{c} \Delta_{c}+\gamma^{2})+4 \gamma^{2} \omega_{c}^{2}}{(\omega^{2}-\omega_{c} \Delta_{c}+\gamma^{2})+4 \gamma^{2} \omega^{2}}
\end{equation}

\begin{equation}
S_{2out}(\omega)=\frac{1}{4} \frac{(\omega^{2}-\omega_{c} \Delta_{c}+\gamma^{2})+4 \gamma^{2} \Delta_{c}^{2}}{(\omega^{2}-\omega_{c} \Delta_{c}+\gamma^{2})+4 \gamma^{2} \omega^{2}}
\end{equation}

Figure 3 displays the spectrum of squeezing in the dispersive-adiabatic limit (left plot) and the strong-coupling-adiabatic limit (right plot). We note, as expected, that in the strong-coupling regime, squeezing at the cavity output is significantly enhanced compared to that in the dispersive regime. Further in both the limits, increasing $U_{0}$, increases the squeezing. We also observe that in the strong-coupling regime, the $S_{1out}(\omega)$ quadrature is squeezed while in the dispersive limit $S_{2out}(\omega)$ quadrature is squeezed. Another striking difference noticed is that in the dispersive regime, the maximum squeezing is at $\omega=0$ while in the strong-coupling regime the maximum squeezing shifts symmetrically to the both sides of $\omega=0$.

\begin{figure}[h]
\hspace{-0.0cm}
\begin{tabular}{cc}
\includegraphics [scale=0.55]{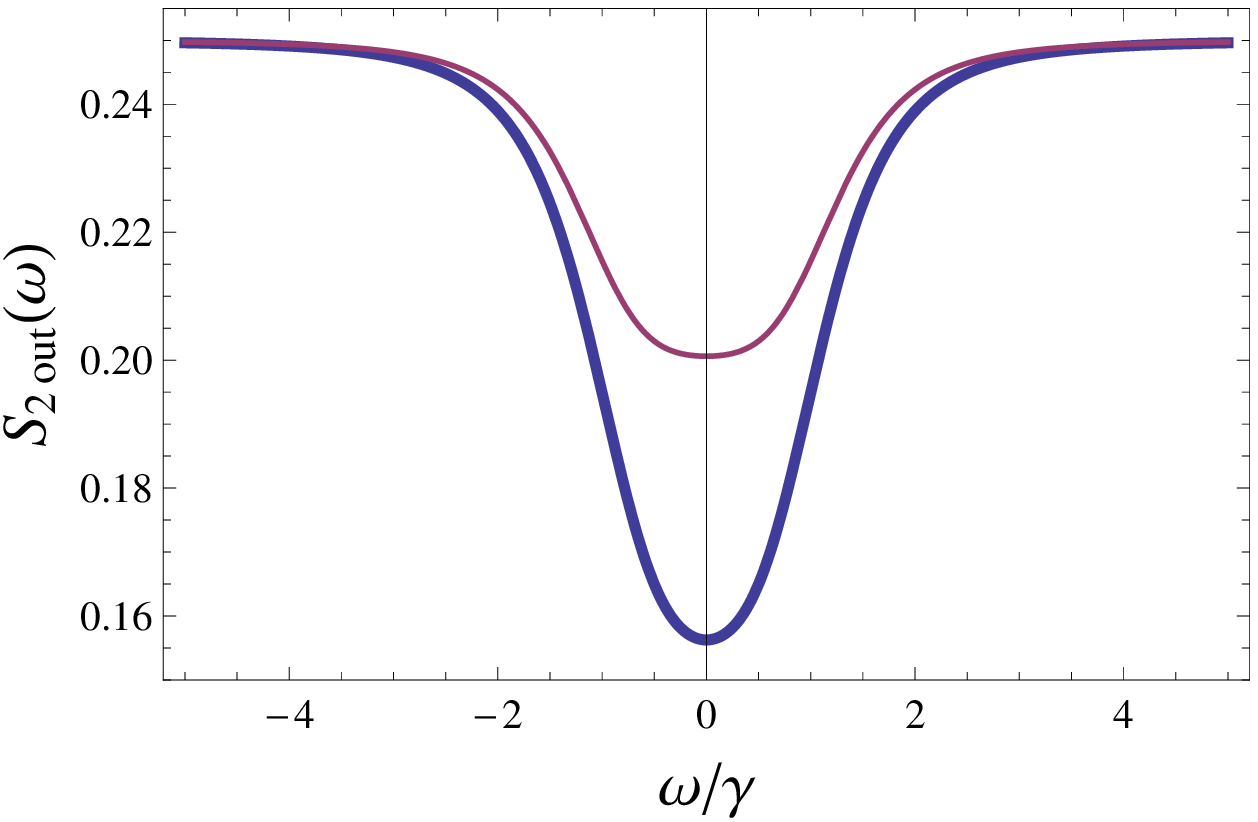} \includegraphics [scale=0.55] {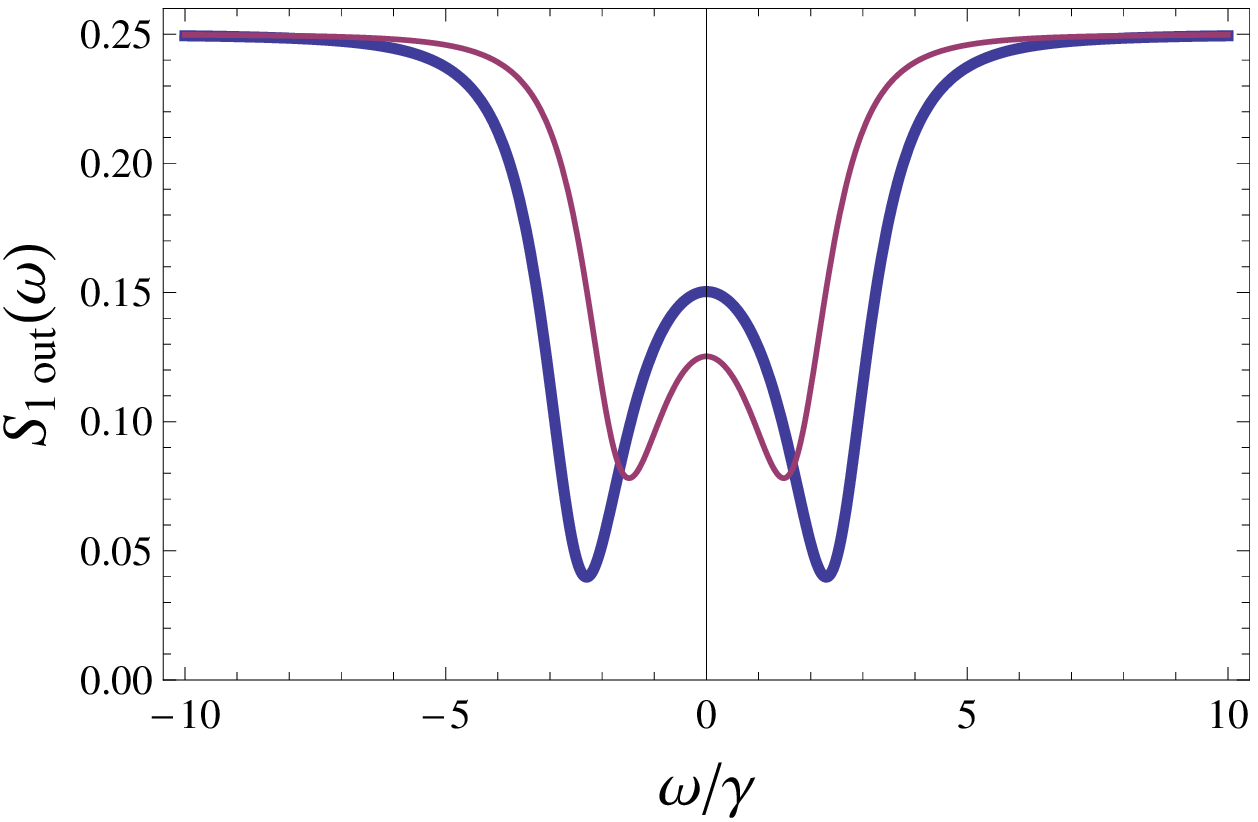}\\
 \end{tabular}
\caption{Left Plot: Spectrum of squeezing in the dispersive-adiabatic limit for $U_{0}=0.01 \gamma$ (thin line) and $U_{0}=0.1 \gamma$ (thick line), $\alpha=0$. Right Plot: Spectrum of squeezing in the strong-coupling-adiabatic limit for $U_{0}=0.5\gamma$ (thin line) and $U_{0}=1.2 \gamma$ (thick line), $\alpha=2.1$}
\label{f3}
\end{figure}

\section{Conclusions}

We have analyzed the potential of a system comprising of a collection of two-level atoms inside an optical cavity, for the preparation of squeezed state of light. In the adiabatic limit the system behaves like a nonlinear medium, which is capable of generating squeezed state of light. In particular, we have compared the results for the dispersive-adiabatic and strong-coupling-adiabatic limit. We have found that squeezing of the output light is significantly higher in the strong-coupling regime. The dispersive limit though easier to implement is only able to squeeze the output light moderately. On the other hand, the dispersive limit was found to be suitable to observe the Dicke superradiant quantum phase transition where the $\vec{A}^{2}$ term is negligible.

\begin{acknowledgments}
A. Bhattacherjee acknowledges financial support from the Department of Science and Technology, New Delhi for financial assistance vide grant SR/S2/LOP-0034/2010.
\end{acknowledgments}

\end{document}